\documentclass{ar-1col-S2O}
\usepackage[numbers]{natbib} % or [authoryear]
\usepackage{graphicx}
\usepackage{url}
\usepackage{amsmath}
\usepackage{mathtools}
\usepackage{amssymb}
\usepackage{subcaption} 
\usepackage{xcolor} 
\usepackage{todonotes}

\begin{document}

% Page header
\markboth{Angheluta et al.}{Topological defects in polar active matter}

% Title
\title{Topological defects in polar active matter}

%Authors, affiliations address.
\author{Luiza Angheluta$^1$, Anna Lång$^2$, Emma Lång$^2$, and Stig Ove Bøe$^2$
\affil{$^1$Njord Centre, Department of Physics, University of Oslo, Oslo, Norway, 0371; email: luizaa@fys.uio.no}
\affil{$^2$Department of Microbiology, Oslo University Hospital, Oslo, Norway, 0372; email: s.o.boe@medisin.uio.no}}

%Abstract
\begin{abstract}
Polar active matter - including animal herds, aggregates of motile cells and active colloids - often forms coordinated migration patterns, such as flocking. This orderly motion can be disrupted by full-integer topological defects representing localized disturbances where polar alignment is lost. Such polar defects can serve as key organizing centers across scales, sustaining collective behavior, such as swirling motion and other large-scale coherent states. While significant progress in understanding active matter principles have been made in recent years, a quantitative understanding of how topological defects influence active polar matter is needed. We present a brief overview of recent experimental observations in synthetic active colloids and various biological systems. We describe how polar defects mediate dynamical transitions and contribute to the spontaneous emergence of large-scale coherent states. We also discuss theoretical advancements in physical modeling of coupled processes involving polar defects and collective behavior in active polar matter.
\end{abstract}

%Keywords, etc.
\maketitle

%Table of Contents
\tableofcontents

% Heading 1
\section{Introduction}
Polar active matter is a conceptual umbrella for diverse systems composed of interacting particles with inherent directional motility, such as migrating animals,  motile cell collectives, and self-propelled colloids. Unlike passive systems, such as ferromagnets, liquid crystals, or superfluids, polar active matter consists of particles that consume energy to propel themselves while also interacting with each other and their environment. These systems are captivating with their ability to spontaneously form large-scale spatio-temporal patterns of orchestrated dynamical processes. 

A central mathematical object for describing this coordinated motion is the vector order parameter field, a metric quantifying alignment, which has zero magnitude in a disordered state (e.g. uncoordinated motion) and reaches its maximum value in a fully coordinated state (e.g. flock state). As in other symmetry-broken states, polar order is often disrupted by topological defects, representing localized tears where order breaks down. 
Fundamental polar defects, including vortices, asters, spirals, and anti-vortices, act as key organizing centers, bridging microscopic interactions and macroscopic behavior. They drive self-organization across scales and mediate dynamical transitions in polar active systems.

While topological defects are well understood in passive condensed matter systems, their role in active matter remains an active area of research. More attention has recently been given to understanding the role of topological defects in polar active systems. In bacterial colonies, polar defects regulate collective motility and phase separation. In animal tissues, polar defects drive collective cell migration and function as organizing hubs for tissue patterning during development and homeostasis. In synthetic active colloids, such as Quincke rollers, polar defects sustain large-scale vortical patterns. 

This review explores recent advances in the study of polar ordering and topological defects in polar active matter. By examining the interplay between activity, defect dynamics, and self-organization, we provide a comprehensive perspective on how topological defects mediate emergent dynamic behavior in biological and synthetic active polar systems.

%------ intro section of defects------
\section{Full-integer topological defects}
\begin{figure}
    \centering
    \includegraphics[width=\linewidth]{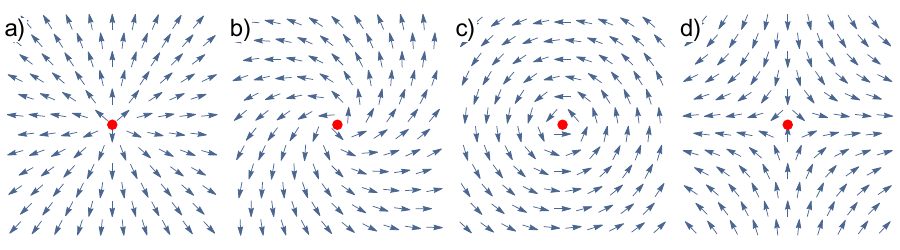}
    \caption{Spatial configuration of orientation field, $\theta$, induced by polar defects: a) aster; b) spiral; c) vortex; d) $-1$ defect.}
    \label{fig:defect_type}
\end{figure}

\begin{figure}
    \centering
    \includegraphics[width=\linewidth]{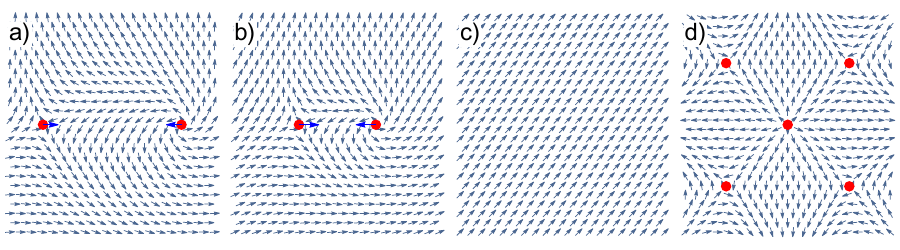}
    \caption{(a-c) \textbf{Uniform order through defect pair annihilation}: one $+1$ defect and one $-1$ defect attract each other and eventually disappear as a pair; d) \textbf{Defect order}: stable spatial configuration of asters and $-1$ defects promoting persistent order, as discussed in Refs.~\cite{mondal2025coarsening,laang2024topology}.}
    \label{fig:defect_order}
\end{figure}

Imagine you are observing a flock of birds flying in unison. Their uniform flying direction represents an \emph{ordered state}, with each bird’s motion represented by an arrow pointing in the same direction. A topological defect, while a mathematical concept at its core, is a tangible entity that can be directly observed through the distinctive spatial pattern it imprints on the surrounding order (Figure~\ref{fig:defect_type}). A swirly motion disrupts the uniform flock such that birds follow spiral trajectories. This distortion pattern is induced by a spiral defect. Similarly, a vortex defect corresponds to a rotating pattern whereby particles orbit around the defect center. These coherent patterns induced by topological defects feed into the emergence of large-scale behaviors in both synthetic and biological active matter systems.
When multiple defects co-exist, they interact with each other as quasi-particles through long-range forces. They can annihilate or nucleate as dipoles, i.e. pairs of defects and  anti-defects (Figure~\ref{fig:defect_order}a-c), and can form complex spatio-temporal patterns with underlying coherency despite their chaotic appearance (Figure~\ref{fig:defect_order}d).

Mathematically, topological defects are singularities where global order is locally broken in such a way that it cannot be patched by any smooth deformation. They are classified by their topological charges, determined by winding numbers tied to order parameters. Order and topology help us understand different states of matter and associated phase transitions in a generic way, i.e. not specific system. A well-known example of this is the Kosterlitz-Thouless transition for two-dimensional (2d) systems with rotation, translation or gauge symmetries~\cite{kosterlitz1973ordering, kosterlitz2016kosterlitz}. 
Ordered systems that break underlying continuous symmetries undergo order$\leftrightarrow$disorder phase transitions, which are mediated by topological defects. For instance, the melting of crystals, liquid crystals or superfluids is associated with unbinding of defect dipoles and the destruction of long-range order. Similarly, the phase-ordering kinetics from a high-temperature quench is mediated by a gradual annihilation of defect dipoles leading to the formation of quasi-long range order. 

It turns out that the defect-mediated transition is a versatile theoretical framework that applies to dynamical transitions in active matter systems beyond the Kosterlitz-Thouless theory. In active nematics, this has been applied to study isotropic-nematic transition  and the transition to active turbulence as function of activity, e.g. Refs. ~\cite{doostmohammadi2018active,shankar2018defect}. Interestingly, only in the last few years, more attention has been dedicated to studying the role of full-integer defects in mediating the formation of persistent order or dynamical transitions in polar active matter. 

On general terms, collective order associated with rotational symmetry can be described by a vector field, i.e. the polarization $\mathbf P = \|\mathbf P\|\left[\cos(\theta)\mathbf e_x+\sin(\theta)\mathbf e_y\right]$, where $\mathbf e_x,\mathbf e_y$ are unit vectors along $x$ and $y$ axes. For superfluids, the order parameter is the scalar complex field corresponding to the superfluid wavefunction $\psi = \|\psi\| e^{i\theta}$. Given the correspondence between complex and vector representations in 2d, the topological defects in $\psi$ and $P$ are mathematically equivalent~\cite{skogvoll2023unified}, and defined by pointwise singularities in the orientational field $\theta(\mathbf r)$. Any contour integral enclosing a phase singularity at a given position $\mathbf r_0$ picks up a net phase shift 
\begin{equation}
    \oint\limits_{C_0} d\theta = \oint\limits_{C_0}  d\mathbf l\cdot\nabla \theta = 2\pi q,
\end{equation}
with the winding number $q\in \mathbb{Z}$ defining the topological charge. From Stokes theorem, it follows that the phase gradient is curl-free except for $\mathbf r_0$, such that 
\begin{equation}
    \epsilon_{ij}\partial_i\partial_i \theta = 2\pi q \delta(\mathbf r-\mathbf r_0).
\end{equation}
The corresponding solution of the phase gradient for $q = \pm 1$ is the irrotational vortex $\nabla \theta = \frac{q}{2\pi r} \mathbf e_\phi$, where $r$ and $\phi$ are the polar coordinates relative to the defect position and $\mathbf e_\phi  = -\sin(\phi)\mathbf e_x+\cos(\phi)\mathbf e_y$. The phase gradient represents the superfluid velocity, which is indirectly observed by measuring of the superfluid current (momentum density), $\mathbf J = \Im(\psi^*\nabla\psi)= \|\psi\|^2 \nabla\theta$. The superfluid density regularizes the $1/r$ singularity of the irrotational vortex, by introducing a smooth core region where the superfluid density rapidly vanishes (condensate melting). The actual profile of the vortex core is energetically determined. In the mean-field Gross-Pitaevskii theory, the vortex core is isotropic and approximated by $|\psi| \approx \frac{\Lambda r}{\sqrt{r^2+\Lambda^2}}$, where $\Lambda$ is a constant determined numerically. Thus, the vortical superfluid current $\mathbf J$ is irrotational outside the vortex core (far-field) and vanishes linearly at the phase singularity (near-field). By taking the curl of the superfluid current, $D = \frac{1}{2} \epsilon_{ij}\partial_i J_j=\frac{1}{2} \epsilon_{ij}\Im\left[(\partial_i \psi^*)(\partial_j\psi)\right]$, we obtain a smooth measure of superfluid vorticity, which is zero outside the vortex core and non-zero in the core region, such its sign corresponds to the topological charge $\textrm{sgn}(D(\mathbf r_0))=q$~\cite{ronning2023precursory}. This $D$-field is a smooth defect density field that is globally conserved with the associated defect current density determined by the dynamics of the superfluid wavefunction $\psi$~\cite{skogvoll2023unified}. As a topological quantity, the $D$-field is related to topological invariants, such as the Euler characteristics $\chi$, when integrated over the entire domain $S$~\cite{ronning2023precursory}
\begin{equation}
    \int\limits_S  D d\mathbf r = \frac{1}{2}\oint\limits_{\partial S}  \|\psi\|^2 \nabla \theta \cdot d\mathbf l = \pi \sum\limits_{\alpha\in S} q_\alpha = \chi,
\end{equation}
where the sum is over all vortices inside the spatial domain $S$ and where the superfluid density $\|\psi\|^2=1$ is uniform in the far-field of  defects. 
%----- begin figure -----------

%---- end figure---------

 This formalism can be readily applied to polar systems with the polarization $\mathbf P$ as the fundamental order parameter determined by coarse-graining microscopic polarities, e.g. spins, velocities or other uniaxial polarities. In such systems, the phase gradient of the irrotational vortex is integrated to obtain the orientational field 
\begin{equation}
    \theta = q \phi+\theta_0, 
\end{equation}
were $\phi$ is the angular coordinate of a point relative to the defect position, while $\theta_0$ as the integration constant corresponds to the orientation of the uniform polarization. Notice that $\theta_0$ is not an observable for superfluids, since the superfluid current depends on phase gradients. However, $\theta_0$ is an important intrinsic phase for spatial configuration of $\mathbf P$  induced by polar defects, particularly for $q=+1$ topological defects.  Figure~\ref{fig:defect_type} illustrates corresponding spatial configurations for each type of full-integer defect. Namely $\theta_0 = \pi/2$ corresponds to a vortex configuration, i.e. $\mathbf P = \|\mathbf P\| \mathbf e_\phi$;  $\theta_0 = 0 (\pi)$ for an outward (inward) aster, i.e. $\mathbf P = \pm\|\mathbf P\| \mathbf e_\phi^\perp$; $\theta_0 \in (0, \pi/2)$ is an outward spiral and $\theta_0 \in (\pi/2,\pi)$ is an inward spiral~\cite{ronning2023spontaneous}. For $q=-1$ defects, $\theta_0$ does not change the profile of $\mathbf P $. The full-integer defects are also referred to as polar defects. 

Based on Halperin-Mazenko formalism, a recent non-singular defect field theory has been proposed as a generic formalism for tracking the location and kinematics of topological defects as moving zeros in order parameters~\cite{skogvoll2023unified}. For the polarization field $\mathbf P$, the corresponding $D$-field is given by~\cite{skogvoll2023unified,andersen2023symmetry}
\begin{equation}
    D = \frac{1}{2}\epsilon_{ij}\epsilon^{kl}(\partial_iP_k)(\partial_j P_l).
\end{equation}
Due to its connection with topological defects, the $D$-field is a conserved field and its associated defect current~\cite{skogvoll2023unified,andersen2023symmetry}
\begin{equation}
    J^{(D)}_i = -\frac{1}{2}\epsilon_{ij}\epsilon^{kl}(\partial_tP_k)(\partial_j P_l),
\end{equation}
determines the motion of defects as well as their creation and annihilation~\cite{skogvoll2023unified}. This defect field theory can be applied to other systems with broken continuous symmetries, including crystals~\cite{skaugen2018dislocation,skogvoll2023unified} and quasicrystals~\cite{de2024mesoscale}. 

%----- begin figure -----------

%---- end figure---------

%----- milling states ------
\section{Milling states from alignment interactions}

\begin{figure}
    \centering
    \includegraphics[width=\linewidth]{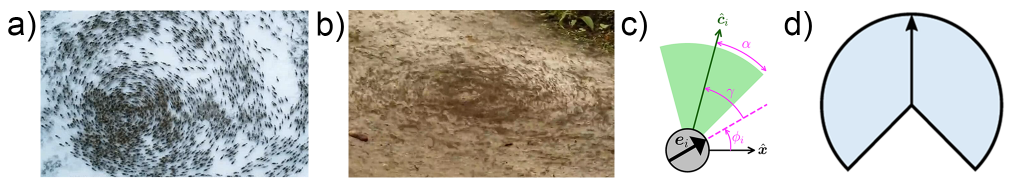}
    \caption{(a-b) Examples of milling states in a reindeer herd and an ant colony (adapted from reference ~\cite{zhou2023circular}); c) Sketch of perception-dependent motility (adapted from reference~\cite{saavedra2024swirling}); d) Sketch of field of view with a blind angle (adapted from reference~\cite{costanzo2018spontaneous}).}
    \label{fig:milling}
\end{figure}

The Vicsek model~\cite{vicsek1995novel} describes flocking behavior in animal systems, from insect swarms to bird flocks and mammal herds, based on the emergence of uniformly coordinated motion through alignment interactions. This occurs when each individual aligns its movement direction with the average direction of its neighbors within a certain radial distance~\cite{couzin2002collective}. 
By model construction, this orientational alignment promotes uniform translational motion. Thus, rotational motion is not manifested spontaneously and instead appears due to geometric constraints through alignment with the confining boundaries (anchoring). However,  
many animal flocks, including ants~\cite{schneirla1944unique}, worms~\cite{franks2016social}, reindeer~\cite{espmark2002behavioural} and fish~\cite{couzin2002collective,harvey1999putative}, as well as active colloids, such as active Brownians with time-delayed interactions \cite{wang2023spontaneous,costanzo2022effect}, also exhibit spontaneously formed milling states in open (unconfined) space (Figure \ref{fig:milling}a-b). 

%----- begin figure -----------

%---- end figure---------

Modified Vicsek models have been formulated \cite{costanzo2018spontaneous,lavergne2019group, costanzo2022effect, saavedra2024swirling} to allow for the spontaneous formation of milling states with a solid-body rotation $\mathbf v(\mathbf r) \sim \Omega \mathbf r^\perp$, where $\mathbf r^\perp = [-y, x]$. We highlight two such recent model formulations. Let us consider a set of self-propelled particles labeled by index $i = 1,\cdots N$ with speeds $v_i$ and $\mathbf e(\theta_i) = [\cos(\theta_i),\sin(\theta_i)]$ the unit vectors determined by the orientations of motion $\theta_i$. In the original Vicsek model, all particles have the same motility $v_i = v_0$.

\subsubsection*{Perception-dependent motility} The overdamped motion is given by \cite{saavedra2024swirling}
\begin{eqnarray}
     \dot{\mathbf r}_i &=& v_i \mathbf e(\theta_i) +\mathbf F_i\\
     \dot{\theta}_i &=&\sqrt{2D_r}\eta_i,
\end{eqnarray}
where $\eta_i\in[0,2\pi]$ is the orientational white noise with the rotational diffusivity constant $D_r$, and $\mathbf F_i$ is a gradient force determined by a pairwise attraction-repulsion potential. Particle $i$ has its own self-propulsion speed $v_i$ determined by restricted field of view, or the perception cone $c_i$, through cone parameters such as the cone half-width angle $\alpha$ and the  misalignment angle $\gamma$ of the direction of the cone's symmetry axis with respect to the self-propulsion direction $\mathbf e_i = [\cos(\theta_i),\sin (\theta_i)]$ (Figure \ref{fig:milling}c). The perception cone restricts the set of interacting neighbors through this function
\begin{equation*}
    P = \sum_{j\in c_i} \frac{1}{r_{ij}}, \quad \textrm{ if } r_{ij}<r_c
\end{equation*}
where $r_c$ is the maximum perception distance. The homogeneous perception $P_0 = \alpha \rho R_0$ is attained when the particle is in the center of a circular region of radius $R_0 = r_c/2$ and with homogeneous density $\rho = N/(\pi R_0^2)$. Then, the self-propulsion speed is modified such that 
\begin{equation}
     v_i = 
\begin{cases}
       v _0,& q_i>q^* \\
        0,& \mbox{otherwise}
    \end{cases}
\end{equation}
where $q_i = P/P_0$ is the normalized perception and $q^*$ is the perception threshold. Depending on the misorientation angle $\gamma$ and threshold $q^*$, different vortex profiles are observed from a compact non-swirling ($\gamma=0$) to a solid-body rotation ($\gamma\ne 0$) and intermediate states with clusters that have a fluid-like core with a rotating outer layer or a core of solid body rotation driven by the outer layer activity~\cite{saavedra2024swirling}.   

\subsubsection*{Field of view with a blind angle}
The equation of motion in discrete time with increment $\Delta t$ reads as~\cite{costanzo2018spontaneous,costanzo2022effect}, 
\begin{eqnarray}
 \mathbf r_i(t+\Delta t) &=& \mathbf r_i(t)+ v_0 \mathbf e(\theta_i) \Delta t\\
 \theta_i(t+\Delta t) &=& 
\begin{cases}
\langle \theta_j (t-\tau)\rangle_{r,\phi} +\Delta \theta_i, & |\Delta\Theta_i|<\Omega\\
\theta_i(t)+\Omega +\Delta \theta_i, & \Delta\Theta_i\ge \Omega\\
\theta_i(t)-\Omega +\Delta \theta_i, & \Delta\Theta_i\le -\Omega\\
\end{cases}
\end{eqnarray}
where $\Delta\theta_i$ is the rotational noise which is uniformly distributed in the interval $[-\eta/2,\eta/2]$, where $\eta$ is the noise parameter. 

In this model, particles can reorient their self-propulsion direction to the average direction of their neighbors within an interaction range of radius $r$ and field of view $\phi$, which is denoted by $\langle \theta_j (t)\rangle_{r,\phi}$ (Figure \ref{fig:milling}d). $\Delta\Theta_i = \theta_i(t)-\langle \theta_j (t-\tau)\rangle_{r,\phi}$ is the difference between the current orientation of particle $i$ and the average orientation of its interacting neighbors. When the average self-propulsion direction falls in the blind zone, the particle rotates its self-propulsion direction by $\Omega\in [0,\pi]$, so that the average direction is within the particle's field of view. Non-zero $\tau$ corresponds to time-delayed alignment interactions.  

%---------------------------
\section{Defects in active polar fluids}
\subsection{Active colloids: experiments}
Colloids are small particles suspended in a fluid, typically ranging from a few nanometers to several micrometers in size. Microswimmers, which are colloidal systems capable of self-propulsion, have become significant model systems for examining polar active matter. 

A notable experimental system involves suspensions of Quincke rollers, which remain stationary until activated by a uniform electric field through a mechanism that induces autonomous rolling in random directions, also known as Quincke rotation. In dilute suspensions, these rollers undergo a flocking transition, aligning into a coherent flow along a preferred direction under the influence of a DC electric field~\cite{bricard2013emergence}. When confined within a disk-shaped geometry, they form a large-scale vortical flow that spans the entire system (several millimeters in diameter)~\cite{bricard2015emergent, chardac2021topology}. 

Experiments on dense active colloids show that velocity alignment is strongly coupled with density fluctuations, leading to the emergence of complex spatio-temporal patterns~\cite{chardac2021topology}. A similar phenomenon has been observed in actin filament networks, where above a critical density, individual movements transition into a collective, defect-mediated phase~\cite{schaller2013topological}. Due to the coupling of velocity and density fluctuations (compressible flows), the coarsening dynamics is facilitated by a network of domain walls interacting with polar defects~\cite{geyer2018sounds}. The coupling of self-advection and density gradients leads to the formation of interconnected, elongated density patterns resembling bow ties, whereby a $-1$ defect in the flow field is surrounded by four domain walls~\cite{chardac2021topology}. 

Using the same experimental system based on Quincke rollers in circular confinement, it was also shown that the strong coupling between environmental disorder and self-propulsion can lead to pinned vortices, and the transition to a dynamic vortex-glass state characterized by correlated flows sustained by pinned defects ~\cite{chardac2021emergence}. The flocking transition in the presence of disorder has been theoretically studied for incompressible polar fluids, e.g. Refs.~\cite{toner2018hydrodynamic,zinati2022dense,chen2022hydrodynamic}. Despite these advances, fundamental questions about the universality of these large fluctuations and defect-mediated transitions in active colloids remain open for future studies. 

%-------------------------------------------
\subsection{Compressible Toner-Tu fluids}
Toner and Tu proposed a minimal hydrodynamic model of flocking behavior of self-propelled particles as an active polar fluid ~\cite{toner1998flocks}. The order parameter for the polarized migration is given by the coarse-grained velocity field $\mathbf v$. The model contains the mass conservation for the particle density $\rho$ coupled with the momentum balance for the velocity field $\mathbf v$ given by ~\cite{chardac2021topology,mondal2025coarsening}
\begin{eqnarray}
  \partial_t \rho +\nabla\cdot(\mathbf v \rho) &=&0 \\
  \partial_t \mathbf v +\lambda \mathbf v\cdot \nabla \mathbf v &=& (\alpha-\beta |\mathbf v|^2)\mathbf v+ D\nabla^2 \mathbf v-\sigma \nabla \rho,\label{eq:v}
\end{eqnarray}
 where $\lambda$ is the self-propulsion parameter that breaks the Galilean invariance for $\lambda \ne 1$. Additional parameters are $\alpha$ and $\beta$ that determine the magnitude of local order, i.e. the magnitude of the uniform flow in steady-state is $v_0 = \sqrt{\alpha/\beta}$. The kinematic viscosity is denoted by $D$, and $\sigma$ is the coupling parameter for the feedback of density variation in the velocity field. Notice that for $\lambda = \sigma =0 $, Eq.~(\ref{eq:v}) reduces to the classical Ginzburg-Landau equation. In the classical Ginzburg-Landau theory, the phase-ordering kinetics from an initial quench is driven by the pair annihilation of $\pm 1$ defects in the order parameter field (Figure~\ref{fig:defect_order}a-c), while any initially formed domain walls are quickly smeared out by diffusion. However, for polar flocks, the coupling between migration velocity and density gradients can lead to stable domain walls co-existing and interacting with polar defects~\cite{chardac2021topology,mondal2025coarsening}. 
\subsubsection*{Vortex profile}
In Ref.~\cite{chardac2021topology}, the steady-state  and long-wavelength limits of Eq.~\ref{eq:v}, 
\begin{equation}
    \lambda \mathbf v\cdot \nabla \mathbf v +\sigma \nabla \rho =0,
\end{equation}
is solved analytically using an ansatz solution for density-dependent velocity corresponding to a stationary vortex. 
The near- and far-field asymptotic expansion of the analytical solution of the density profile reads as, 
\begin{equation}
    \rho \approx \rho_c
    \begin{cases}
    1+\left(\frac{r}{a}\right)^\Lambda, & r\approx a\\
    1+\Lambda\log\left(\frac{r}{a}\right), & r\gg a\\
    \end{cases}
\end{equation}
where $\rho_c$ is the critical density for flocking transition, $\Lambda = \lambda v_0^2/(\sigma\rho_c)$ and $a$ is the vortex core size. This topological state exits above the flocking transition $\rho>\rho_c$ with a compressible velocity field given by \cite{chardac2021topology}
\begin{equation}
    \mathbf v = v_0 \sqrt{1-\frac{\rho_c}{\rho}}\mathbf e_\phi.
\end{equation}
Unlike the irrotational vortex, which decays as $1/r$ away from the vortex, this vortical flow increases with $r$ and
approaches logarithmically the constant-speed vortex, where particles are rotating at the same speed independent of the distance to the vortex core. The stability of this vortex solution needs to be further studied. A recent numerical study~\cite{mondal2025coarsening} shows that the aster-like defects form instead spontaneously and drive anomalous phase-ordering kinetics.

\subsubsection*{Defects kinematics}
The dynamics of polar defects in the velocity field  follows a kinematic equation~\cite{chardac2021topology} 
\begin{equation}
    \gamma_i\dot{\mathbf R}_i = \sum_{j} F^{(C)}_{ij}+\mathbf F_i^{(M)},
\end{equation}
driven by Coulomb-like forces between defects $\mathbf F^{(C)}_{ij}\sim q_i q_j\frac{\mathbf R_i-\mathbf R_j}{|\mathbf R_i-\mathbf R_j|^2}$ and a Magnus-like force 
\begin{equation}
    \mathbf F_i^{(M)} \approx (\dot{\mathbf R}_i^\perp -\mathbf V^\perp_{far} ),
\end{equation}
where $\mathbf F^\perp = [-F_y, F_x]$ and $\mathbf V_{far}$ is the velocity field from the far-field. This Magnus force appears from the self-advection coupled with density gradients and contributes to rotational motion of defects. 

\subsubsection*{Phase-ordering kinematics} The phase-ordering kinetics is driven by the interactions between domain walls and topological defects through pair annihilation of defects with opposite topological charges. The mean density of defects $\rho_d(t)\sim t^{-\alpha}$ decays algebraically with time with a power-law exponent $\alpha$ determined by defect kinematics. In the limit of dominating Coulomb-like forces, the exponent is $\alpha=1$ and is associated with correlation length that increases diffusely with time. Deviations from this scaling law suggests that Magnus forces are also important in the defect annihilation process, and that the interactions between domain walls and defects may also drive the coarsening process. A numerical study of a similar compressible Toner-Tu model shows that the defect coarsening dynamics is anomalously slow on long timescales as the total number of defects saturates to a non-zero value corresponding to a stable configuration of asters separated by $-1$ defects~\cite{mondal2025coarsening} (Figure~\ref{fig:defect_order}d).  

%-----
\subsection{Hydrodynamics of polar active liquid crystals}
An alternative hydrodynamic model of polar active matter has been proposed in Ref.~\cite{amiri2022unifying} based on active liquid crystals.  Active liquid crystals are composed of elongated particles, with rod-like shapes, suspended in a fluid. As active particles, they consume energy to generate translational motion. These systems are characterized by a polarization field $\mathbf p$, representing the coarse-grained alignment of microscopic polarities inherent to head-tail elongated shape of individual particles. The elongated suspensions can generate fluid flows through disturbances in the polarization field. The evolution of $\mathbf p$ is governed by the relaxation of a free energy functional 
\begin{equation}
    \mathbf F = \int d\mathbf r \left\{A \left(\frac{p^4}{4}-\frac{p^2}{2}\right)\right\}+\frac{K_p}{2}|\nabla\mathbf p|^2+\frac{K}{2}\left|\nabla \cdot \left(\mathbf p\mathbf p^T-\frac{p^2}{2}\mathbf I\right)\right|^2,
\end{equation}
which allows for polar alignment controlled by the stiffness parameter $K_p$, and apolar (nematic) alignment controlled by the stiffness parameter $K$. The dynamical equations for the evolution of the polarization coupled with an incompressible flow velocity $\partial_i u_i=0$ read as 
\begin{eqnarray}
    \rho_0\partial_t u_i &=& \partial_i \sigma_{ij}+\alpha_p p_i-\xi u_i\\
    \frac{D p_i}{Dt}-(\lambda E_{ij}+\Omega_{ij})p_j &=& -\frac{1}{\gamma}h_i,
\end{eqnarray}
where $h_i = -\delta \mathcal F/\delta p_i$ is the molecular field driving the relaxational dynamics, and $\xi$ is the friction coefficient for frictional drag with a substrate. The flow alignment with the strain rate $E_{ij}$ is controlled by $\lambda$, and $\Omega_{ij}$ is the vorticity tensor. The stress $\sigma_{ij} = \sigma_{ij}^{vis.}+\sigma_{ij}^{pas.}+\sigma_{ij}^{act.}$is composed of viscous stress $\sigma_{ij}^{vis.} = 2\eta E_{ij}$, passive stress $\sigma_{ij}^{pas.} = -P\delta_{ij}+C_{ijkl}p_k h_l$, and active stress $\sigma_{ij}^{act.} = -\zeta Q_{ij}$ determined by the nematic order parameter $Q_{ij} = \left(p_ip_j-\frac{p^2}{2}\delta_{ij}\right)$. Thus, the model includes two sources of activity: i) crawling (polar) force due to self-propulsion controlled by parameter $\alpha_p$; and ii) dipolar force due to active stress controlled by parameter $\zeta$. When both polar and apolar interactions are finite and non-zero, this minimal model predicts the emergence of an active turbulent state characterized by vortices and jets in the flow velocity induced by the co-existence of  half-integer and full-integer topological defects in the polarization field~\cite{amiri2022unifying}. When only polar interactions are allowed, i.e. $K=0$ and with increasing the activity parameter $\zeta$, the system transitions from a defect-free active turbulence to an active turbulence laden with full-integer defects in the polarization field~\cite{andersen2023symmetry}. At the onset of defect-laden turbulence, the kink walls in the polarity field become unstable to pair nucleation of polar defects with opposite charge that subsequently dissociate.

Polar and dipolar forces induce distinct incompressible flow patterns around fixed polar defects~\cite{ronning2023spontaneous}. 
The incompressible flow pattern induced by dipolar forces for an isolated $\pm 1$ defect has a far-field radial decay determined by the hydrodynamic dissipation length $l_d = \sqrt{\eta/\xi}$. For any $+1$ defect (i.e. any $\theta_0$ distinguishing between an aster, a spiral or a vortex), the incompressibility condition removes the radial flow component leading to a purely vortical flow~\cite{ronning2023spontaneous}. The dipolar flow induced by a $-1$ defect has instead an 8-fold azimuthal symmetry. By contrast, the crawling force $\alpha_p \ne 0$ as a measure of self-propulsion, induces also a purely vortical flow pattern for a $+1$ defect (except the perfect aster $\theta_0=0$), but with the far-field solution~\cite{ronning2023spontaneous}
\begin{equation}
    \mathbf v =  \sin(\theta_0) \left(1-\frac{1}{r^2}\right)\mathbf e_\phi, \qquad r\gg 1
\end{equation}
that saturates at large distances to the far-field speed, similar with the vortex solution in the compressible Toner-Tu model. The polar flow profile around a $-1$ defect has a 4-fold azimuthal symmetry. This polar force induces a net torque force acting on the $\pm 1$ defects and promotes polar ordering through defect pair annihilation.  

Compressibility effects have been recently explored in Ref.~\cite{zhao2025integer}, where it is shown that the coupling of the active dipolar flow to cell density gradients generates inward spiral flows, thus explaining that accumulation of cells at the center of patterned $+1$ defects (for any $\theta_0$) as observed experimentally for neural progenitor cell
monolayers~\cite{zhao2025integer} or fibroblast monolayers~\cite{endresen2021topological}.

%--------polar solids ------
\section{Polar defects in animal tissues}
\subsection{Epithelial migration}

Epithelial tissue covers all surfaces and cavities in multicellular organisms. The cells in these tissues are tightly connected, forming single or multi-layered sheets that regulate transport and provide protection against the external environment. In adult organisms, these tissues typically exist in a resting state sustaining controlled cellular turnover through homeostatic processes. This stable state contrasts sharply with their behavior during developmental morphogenesis or wound healing, when epithelia undergo a dynamic transition to coordinated collective cell migration. 
The transition of epithelial tissues from an immobile state during homeostasis to a state characterized by extensive collective cell migration, poses several major challenges: cells must be able to unjam from a tightly packed, mechanically constrained configuration and simultaneously self-organize into coherent, directionally aligned collective motion, while maintaining cell-to-cell adhesion. During episodes such as epidermal wound healing or embryo development, these migratory behaviors manifest as moving epithelial sheets that span several millimeters in length, comprising hundreds of thousands of epithelial cells that collectively migrate in a unified direction across curved surfaces~\cite{laang2018coordinated, malinverno2017endocytic, solnica2012gastrulation, park2017tissue}. The result is a coherent, gliding "carpet" of cells, where each self-propelling unit coordinates with neighbors through mechanical and chemical signaling, translating local forces into large-scale, directional motion.

Multiple mechanisms have been proposed to underlie the unjamming transition that facilitates collective motility in epithelial monolayers. In the context of epithelial wound healing, migration toward the wound edge is orchestrated by specialized leader cells, which guide directional movement while propagating mechanical and biochemical signals to adjacent cells at the back. These signals promote unjamming in regions distal to the leading edge, thereby fluidizing epithelial layers beyond the immediate wound boundary \cite{trepat2009physical, poujade2007collective, hino2020erk}. 

In confined bronchial epithelial monolayers, the unjammed-to-jammed transition was shown to depend on dynamic cell shape remodeling, where the unjammed state correlated with increased fluctuations in the cell shape index (\textit{p}), a dimensionless parameter reflecting cell geometry \cite{park2015unjamming}. Furthermore, over-expression of the oncogenic GTPase RAB5 in MCF-10A mammary epithelial cells triggered a flocking transition in otherwise jammed monolayers, linking endocytic recycling mediated by RAB5 to tissue-scale unjamming \cite{malinverno2017endocytic}. Additionally, quiescent HaCaT keratinocyte monolayers exhibited large-scale, EGFR-dependent collective migration upon serum re-exposure, suggesting that unjamming represents an intrinsic, latent property of resting epithelial tissues which is primed for activation by growth signals \cite{laang2018coordinated}.

Epithelial cell sheets exhibit a striking capacity for collective rotational motility. Even at the two-cell stage, cells connect to form a unified rotating body, and when confined to disk-shaped micropatterns, self-organizing into sustained rotational motion is observed \cite{doxzen2013guidance, tanner2012coherent, nanba2015cell, segerer2015emergence, samperio2021bacterial}. Under square confinement, HaCaT keratinocytes have been observed to form oscillatory rotational movement \cite{peyret2019sustained}. Moreover, when quiescent HaCaT cells are subjected to re-activation by serum-borne growth factors under circular confinement, the formation of an inward-pointing macroscopic spiral, spanning several millimeters in diameter, emerges \cite{laang2022mechanical}. Similarly, Madin-Darby canine kidney (MDCK) epithelial cells plated as a circular pattern form expansive, tissue-spanning vortices \cite{heinrich2020size}. In addition, whole-organ time-lapse imaging of the cochlear isolated from mouse embryos has shown that spiraling +1 defects move along the lateral side of the cochlear duct during development~\cite{ishii2021retrograde}.

Recent studies have demonstrated that epithelial monolayers can self-organize spontaneous collective migration with long-range directional order through a topology-driven flocking transition. These experiments were performed using HaCaT keratinocytes, which represent immortalized cells isolated from the human epidermis. HaCaT cells can be programmed into a static quiescent state by prolonged growth factor depletion. Reactivation of these cell sheets with serum-borne growth factors triggers EGFR-dependent migration, producing velocity-aligned movement across large scales~\cite{laang2018coordinated}. Using live cell microscopy, the transition of HaCaT monolayers was monitored as they evolved from a disorganized static state to an activated state characterized by coordinated collective migration. Enhanced self-propulsion activity after serum activation led to the formation of multiple ±1 defect pairs, which subsequently initiated progressive annihilation, resulting in system-spanning order at millimeter scales, with a single +1 defect persisting at the center of the disk~\cite{laang2022mechanical, laang2024topology}.

The experimental design and phase-ordering kinetics in these epithelial studies bear striking parallels to experiments using colloids activated by rapid rotation \cite{chardac2021topology, chardac2021emergence}. In both systems, a transition from static to motile states drives defect-mediated ordering dynamics, culminating in a single +1 defect at the disk center, consistent with the Euler characteristic ($\chi = 1$) for circular confinement.

Analysis of the role of cell-cell junctions in these experiments revealed that topology-guided velocity patterning operates most effectively in a solid-like state, where persistent cell-cell adhesion enables mechanical coupling across the monolayer. Furthermore, a key feature of this defect-mediated coarsening mechanism is its capacity to reverse the direction of polar order through nucleation and subsequent annihilation of topological defects with new pairing configurations. This defect-driven polarity flipping allows the monolayer to adapt its motility direction in response to cell density fluctuations. The defect-mediated migration  directs collective flow toward wound edges \cite{laang2024topology}. However, the theoretical framework or functional mechanism by which topological defects interact with epithelial edges remains unresolved.

\subsection{Polar defects as organizing centers for tissue patterning}\label{Sec:biology}

Full-integer topological defects are observed in the migratory patterns of diverse epithelial surfaces across multiple organs. A unifying principle emerges in which these defects function as organizing hubs, stabilizing and directing collective migration, symmetry breaking, and tissue remodeling. In the following section, we highlight key examples demonstrating how +1 defects integrate epithelial function with tissue patterning (Figure 4).

\subsubsection*{Cornea epithelium}
A prominent example of an epithelial system that supports formation of a macroscopic vortex is the developing corneal epithelium (Figure 4a). Basal epithelial cells migrate from the limbus (a peripheral stem cell niche) to form a curved epithelial sheet covering the eye’s surface. The limbus acts as a geometric confinement, directing cells inward toward the corneal center resulting in a vortical flow pattern.  In mice, the pattern develops from a disorganized cell mass at postnatal day 1, when their eyes open for the first time, and reaches its fully developed form within 6 weeks~\cite{collinson2002clonal,dua1993corneal,collinson2004corneal,nagasaki2003centripetal,di2015tracing,mohammad2015mechanics,iannaccone2012three,aj1973vortex}.

Several mechanisms have been proposed to drive this coarsening process, including chemical gradients, biomechanical cues, and bioelectric signaling ~\cite{aj1973vortex,lemp1989corneal,foster2014differential,jones1996sympathetic,reid2005wound,maurice1965distribution,buck1985measurement,sharma1989kinetics,dua1996vortex}.
However, more recent studies suggest that the role of guidance signals is less important and that the primary force responsible for this emergent pattern is rooted in self-organization through active matter physics principles \cite{lobo2016self, kostanjevec2024spiral}. By incorporating geometric confinement, corneal curvature and polar alignment of neighboring cells into an \textit{in silico} agent-based active matter model, the authors were able to reproduce the spiral patterns of cell migration observed \textit{in vivo}~\cite{kostanjevec2024spiral}. This theoretical model suggests that multiple $\pm 1$ defects  initially form and spread throughout the corneal epithelium. As the coarsening process progresses, these defects collide and annihilate, ultimately producing a stable +1 defect at the center of the matured cornea. Intriguingly, the underlying neural sensory network aligns with this vortex pattern, reflecting the dynamics of the epithelium ~\cite{kostanjevec2024spiral}. In this scenario, the stable +1 defect at the corneal center functions as an organizing hub, stabilizing radial symmetry, and coordinating collective cell migration to maintain epithelial integrity during tissue remodeling.

%----- begin figure -----------
\begin{figure}
    \centering
    \includegraphics[width=\linewidth]{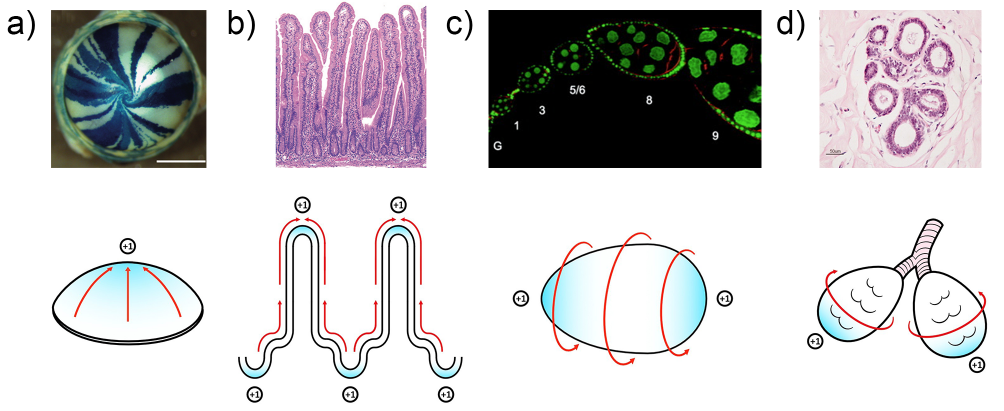}
    \caption{$+1$ defects as centers of tissue organization: a) Murine corneal epithelium (adapted from reference \cite{collinson2002clonal}); b) Human small intestine (adapted from reference \cite{kaye2024normal}); c) \textit{Drosophila} egg chamber development (adapted from reference \cite{horne2005mass}); d) Human mammary gland alveoli (adapted from reference \cite{hassiotou2013anatomy}). Lower panels: illustrations of +1 defects directing the patterning of corresponding tissues (courtesy of Knut Bauer).}
    \label{fig:biological_defects}
\end{figure}
%---- end figure---------

\subsubsection*{Small intestine}
The intestinal epithelium (Figure 4b) is organized into crypts and villi. Villi are finger-like epithelial protrusions extending up to 1 mm in length, and each villus can be surrounded by multiple crypts. Within this system, the center of crypts serves as stem cell compartments, where stem cell division produces new cells. This process generates a migratory stream of differentiating cells from the center of the crypt toward the tip of the villus. Consequently, cell motility results in divergence at the crypt bottom and convergence at the villus tip~\cite{gehart2019tales}. In this context, +1 defects function as hubs for structural shape and function, with those at the crypt bottom acting as centralized stem cell compartments and those at the villus tip serving as sites for cell shedding. The large-scale architecture of the intestinal surface spanning thousands of interconnected villus-crypt units suggests that additional topological features, such as -1 defects, may emerge at transitional regions between converging villi and crypts. However, the presence of -1 defects at the intersection of these structures remain to be confirmed experimentally. 

\subsubsection*{\textit{Drosophila} egg chamber}
During \textit{Drosophila} oogenesis, the egg chamber exhibits rotational collective migration of follicle cells (Figure 4c). These cells, which encapsulate the germline, coordinate movement via actin-based protrusions and planar polarity cues, generating mechanical forces that drive epithelial rotation. This rotation remodels the basement membrane, contributing to the egg’s elongated shape. Due to the ellipsoidal geometry of the egg chamber and topological constraints governed by the Euler characteristic ($\chi = 2$ for a closed ellipsoidal surface), the collective motion generates two $+1$ topological defects, in the form of vortices, at each pole. These defects act as organizing hubs, anchoring rotational forces and symmetry-breaking processes \cite{cetera2014epithelial, cetera2015round}.

\subsubsection*{Mammary gland alveoli}
Mammary gland alveoli are small, sac-like structures within the mammary gland responsible for milk production and secretion (Figure 4d). Recent studies using humanized mammary gland organoids have demonstrated the rotational motion of epithelial cell layers within these tissues. In this experimental approach, cylindrical epithelial branches was observed  transition into spherical alveoli as cells at the branch tips transition from translational motion to persistent collective rotation around the longitudinal axis. This rotational motion generates polar +1 topological defects at the tips, which act as organizational hubs for tension-driven morphogenesis~\cite{fernandez2021surface}.

\subsubsection*{Tumor spheroids}
Rotational flow driven by coordinated cell movements has also been observed in multiple studies using tumor spheroids~\cite{brandstatter2023curvature, chin2018epithelial, cetera2015round,delarue2013mechanical, palamidessi2019unjamming, tanner2012coherent}. These studies illustrate a general concept, where the ellipsoidal or spherical geometry of epithelial surfaces imposes topological constraints that favor the emergence of $+1$ defects.

\subsubsection*{\textit{Hydra} morphogenesis}
In a study of \textit{Hydra} morphogenesis~\cite{maroudas2021topological}, it was demonstrated that topological defects in the nematic order of supra-cellular actin fibers serve as critical organizing centers during \textit{Hydra} regeneration. Unlike polar systems, \textit{ Hydra's} ectodermal actin fibers exhibit nematic alignment, where $+1/2$ defects dynamically fuse to form stable $+1$ defects at the organism's poles. These long-lived $+1$ defects, constrained by the topological requirement of a total charge of +2 on the closed surface, act as mechanical morphogens guiding the emergence of the head and foot. This example underscores that $+1$ defects can act as universal organizing hubs across both polar and nematic systems.

\subsubsection*{Tubulogenesis}
Many epithelial tissues, such as those in the kidney, lung, and glands, exhibit branching and/or tubular morphogenesis. This process is thought to begin with bud formation, where cellular extensions grow perpendicularly from the parental tissue, potentially generating a topological +1 defect in the velocity field at the budding site. Similarly, endothelial tissue undergoes branching morphogenesis during angiogenesis, suggesting that $+1$ defects may also guide sprouting processes and new blood vessel formation. 
A study utilizing the mouse myoblast cell line C2C12 demonstrated tube formation extending from a monolayer confined to a flat disc-shaped surface. After plating, these elongated cells spontaneously self-organized into a spiral configuration with rotational flow, which evolved into spiraling tubes upon pharmacological inhibition of myoblast differentiation \cite{guillamat2022integer}. In line with this, a theoretical study predicted $+1$ polar topological defects as mechanical anchor points that orchestrate three-dimensional (3d) protrusions during epithelial tubulogenesis. These defects, guided by planar cell polarity (PCP) cues, concentrate compressive stresses and act as organizing hubs for out-of-plane structural remodeling. By driving inward cell migration, $+1$ defects induce conical shape formation. Localized tissue fluidization then mediates structural rearrangements at the tip of the cone, facilitating the emergence of tubular structures that protrude from the planar epithelial layer \cite{ho2024role}. Further experimental studies are needed to establish the functional role of $+1$ topological defects in regulating epithelial tubulogenesis.

\subsection{Reconstituted cytoskeletal filaments as experimental active matter systems}

Actin filaments and microtubules are essential cellular components of the cytoskeleton, capable of self-assembling from actin monomers and tubulin dimers into complex dynamic structures. The continuous polymerization and depolymerization of these filaments drive changes in cell shape that are critical for processes such as migration, adhesion, and division. Additionally, these active filaments serve as tracks for motor proteins - such as myosin, kinesin, or dynein - which facilitate intercellular transport ~\cite{cammann2025form, kruse2024actomyosin}.

Motility assays of actin or tubulin with their corresponding motor protein in the presence of ATP, are widely employed for studying the dynamics of active filaments. Depending on experimental conditions, these active filaments can self-organize into collective states with nematic or polar order~\cite{huber2018emergence, roostalu2018determinants}. 
Studies of reconstituted microtubule filaments under conditions that favor polar ordering revealed formation of aster-like defects exhibiting spindle-like spatial organization similar to the microtubuli organization observed in mitotic and meiotic cells  ~\cite{mitchison1984microtubule, ndlec1997self, urrutia1991purified, surrey2001physical}. In confined cylindrical geometries, these asters evolved into vortices as the microtubules elongate and buckle under compressive forces, with microtubule plus ends oriented toward the vortex core~\cite{ndlec1997self}. The same study demonstrated a striking dependency of motility patterns on motor protein concentration. At low motor densities, microtubules self-organized into regular lattices of vortices. Intermediate motor concentrations led to the emergence of aster lattices, while high motor concentrations induced bundling into aligned arrays~\cite{ndlec1997self}. 

Acting filaments in high-density motility assays have revealed collective motion leading to the emergence of $\pm 1$ defects within the velocity field ~\cite{schaller2010polar, schaller2013topological}. Examination of these systems demonstrated that defects sharing the same topological charge can merge to create a  defect with larger topological charge~\cite{schaller2013topological, wollrab2019polarity}, whereas defects with opposite charges undergo pair annihilation ~\cite{schaller2013topological}. The organization of actinomyosin is guided by the coarsening of $+1$ defects, manifesting as asters, which develop into structured contractile networks ~\cite{wollrab2019polarity, soares2011active}. Furthermore, localized polar organization of actin has been detected on a lipid substrate relevant to biological systems ~\cite{sciortino2021pattern}. Recent simulations of the compressible Toner-Tu model show that actin filaments can self-assemble into stable patterns of asters and $-1$ defects that slow down the defect coarsening dynamics~\cite{mondal2025coarsening}.

Beyond 2d self-assembly, contractile  actinomyosin gels can spontaneously form 3d shape - such as domes and saddles - driven by internal stress gradients. Interestingly, these 3d morphologies are influenced more by the initial densities of actin, myosin, and cross-linkers, rather than by confinement geometry or dimensionality~\cite{ideses2018spontaneous, livne2024self}.

The parallels between reconstituted actin systems and the actinomyosin-dependent collective behavior of epithelial cells suggest that \textit{in vitro} assays capture core cytoskeletal mechanisms underlying large-scale tissue dynamics. Further research is needed to understand the interplay between topological defect dynamics in simplified cytoskeletal reconstitutional assays and multicellular coordination driving biological processes like morphogenesis, tissue homeostasis, and wound healing \textit{in vivo}.
%----- begin figure -----------
\begin{figure}
    \centering
    \includegraphics[width=\linewidth]{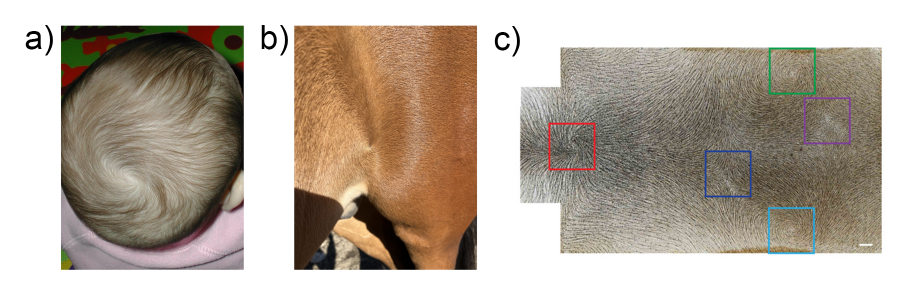}
    \caption{a) Human baby hair whorl (NoJhan via Wikimedia Commons (CC BY-SA 2.5)); b) Hair whorl adjacent the groin of a full-grown horse (courtesy of Pernille Blicher); c) Hair whorls in dorsal skin from Vangl1/2 KO mouse,  postnatal day 7 (adapted from Ref.~\cite{cetera2017planar}).}
    \label{fig:hair_pattern}
\end{figure}
%---- end figure---------

\subsection{Polar defects in animal hair and fur patterns}

The presence of $\pm 1$ defects is commonly observed in the hair and fur patterns of mammals and insects, suggesting that polar ordering governs follicle alignment. In humans, this phenomenon manifests as radially converging patterns on the scalp, forming a central $+1$ defect known as a "whorl" (Figure 5a). Individuals with multiple whorls develop compensatory $-1$ defects to satisfy topological constraints \cite{wunderlich1975hair}. In fully furred animals, numerous ±1 defect can be observed, often evolutionarily conserved and positioned at strategic anatomical locations \cite{lawrence1966gradients, guo2004frizzled6}. For example, horses exhibit $+1$ defects on both sides of the groin area, where their placement facilitates water drainage, keeping this region dry (Figure 5b).

Studies of PCP proteins provide key insights into how mammalian hair forms ordered patterns punctuated by $\pm 1$ defects. Disruption of core PCP genes, such as Frizzled6 and Vangl, interfere with polarity-driven alignment in both epidermal hair follicles and cochlear sensory hair cells \cite{guo2004frizzled6, devenport2008planar, montcouquiol2003identification, wang2005regulation, wang2006order, gubb1982genetic, wong1993tissue} (Figure 5c). In mice with Frizzled6 knockout (KO), analysis at postnatal days 5–8 revealed an increased density of defects compared to wildtype controls. In addition, the investigators discovered a gradual decrease in ±1 defects density during the initial days following birth \cite{wang2010whorls}. This observation aligns with a scenario, wherein Frizzled6 KO mice are born with a higher density of $\pm 1$ defects in their fur compared to their wildtype counterparts, and over time achieve a more ordered state through the dynamics of defect annihilation.

Despite their close association with the epidermis, which represents an active polar system, hair follicles and their patterning align with a static polar system. This is inferred from their origin in epidermal placodes, which represent transient embryonic structures that establish fixed positions and spacing during development. Live imaging of placode formation in mice revealed counter-rotational cell motion driven by PCP signaling, aligning follicles along the anterior-posterior axis \cite{cetera2018counter}. However, the mechanisms governing collective alignment across neighboring follicles remain unresolved. Given that postnatal fur retains fixed defect patterns, this system likely undergoes a dynamic polar ordering phase during early embryogenesis to establish long-range hair follicle alignment.

%--------------------------
\section{Polar defects in self-aligning active matter} 
Alternative flocking mechanisms have been proposed beyond the mutual alignment paradigm of the Vicsek model. Self-aligning active matter is a well-established concept to describe self-organization in dense assembly such as dense biological systems, meta-materials and swarm robotics~\cite{baconnier2024self}. The self-alignment is an intrinsic mechanism whereby active polar particles rotate their polar axis to align with the direction of motion resulting from the forces acting on the particle~\cite{baconnier2024self}. A recent experimental study of interacting Janus colloidal particles (pertaining to the class spherical repulsive active particles)~\cite{das2024flocking} has shown that flocking states can emerge from the presence of a self-aligning torque induced by the asymmetric repulsive force across the particle semi-spheres. 

The generic equations of 2d motion for  interacting polar particles with positions $\mathbf r_i$ and unit vectors $\mathbf e_i$ representing their polar axis are given by~\cite{baconnier2024self}
\begin{eqnarray}
    \dot{\mathbf r}_i &=& v_0 \mathbf e_i+ \frac{1}{\gamma}\mathbf F_i+\sqrt{2D_t} \xi_i\\
    \dot{\mathbf e}_i &=&\beta (\mathbf T_i \times\mathbf e_i) +\sqrt{2D_r} \eta_i \mathbf e^\perp_i,
\end{eqnarray}
where $i= 1,\cdots N$ is the particle label. Gaussian white noise with zero mean and small amplitude corresponds to translational $D_t$ and rotational $D_r$ diffusion. $v_0$ is the self-propulsion magnitude, as a measure of activity with polar symmetry, and $\gamma$ is a frictional damping coefficient. For cell tissues,  $\gamma v_0$ corresponds to the magnitude of the active crawling force. Additional forces 
$\mathbf F_i = \mathbf F_{ext}-\sum_{j\ne i}\nabla_{\mathbf r_i} V_{ij}$ may include external forces (due to a confining potential or other external fields) and gradient forces for attraction/repulsion interactions determined by a pairwise potential $V_{ij}$. The self-aligning torque $\mathbf T = \mathbf e\times \mathbf{w}$, corresponds to the rotation of the polar vector towards the direction of motion, $\mathbf{w} \equiv \dot{\mathbf r}$ (for linear coupling) or $\mathbf{w} \equiv \dot{\mathbf r}/|\dot{\mathbf r}|$ (for non-linear coupling). The parameter $\beta$ dictates how fast the self-propulsion direction rotates towards the direction of the local force. Notice that the non-interacting limit with zero self-aligning torque reduces to the active Brownian particle, e.g. Ref. ~\cite{zottl2023modeling}. Active elastic models can be formulated using nearest-neighbor interactions with spring-like forces on an underlying connectivity graph.  

Although the transition to flocking states as function of packing fraction and model parameters is relatively well studied, there are only few studies exploring topological features. A numerical study~\cite{canavello2024polar} of self-aligning particles with soft interaction potentials (Lenard-Jones type) showed that trapping particles in a harmonic potential (as a proxy to disk confinement) leads to the formation of different orbiting states as a vortex configuration (solid-body rotation), an active crystal - where particles arrange into a lattice that orbits around the harmonic trap, or a polar state formed by small crystalline clusters rotating around the center of the trap. The transitions between these different ordered states depend on the packing fraction and the strength of confinement relative to noise and interaction forces. 

The Euler characteristic of geometric confinement poses global topological constraints, which are often associated with large-scale coherent patterns. Another recent example of this is the emergence of spiral patterns in the collective cell migration in the corneal epithelium~\cite{kostanjevec2024spiral} as discussed in Sec.~\ref{Sec:biology}. Corneal epithelial cells are modeled as soft, self-propelled particles confined as a monolayer on a spherical cap and interacting with their nearest neighbors through spring-like forces. The self-propulsion direction $\mathbf e_i$ can rotate both due to self-aligning torques as well as alignment interactions with $\mathbf e_j$ of nearest neighbors. In addition to active migration and directional alignment, the model includes cell proliferation/extrusion events at rates modulated by the local cell density. The cell migration is confined to the spherical cap through a projection operator that removes the motion normal to the surface. This model reproduces the spontaneous formation of inward spiral flows as observed experimentally in the basal layer~\cite{kostanjevec2024spiral}, also discussed in Sec.~\ref{Sec:biology}. 
    
The phenomenon of collective cell migration guided by topological defects was first experimentally observed in serum-stimulated HaCaT keratinocytes~\cite{laang2024topology}, as discussed in Sec.~\ref{Sec:biology}. The initial phase following stimulation is marked by heightened cell motility coupled with system-wide defect proliferation. Subsequently, the epithelium enters into the second stage, where phase-ordering kinetics is mediated by pairwise annihilation of $\pm 1$ defects such that the large-scale collective migration is dominated by a transient spiral pattern. A similar defect-mediated polar ordering can be predicted by a minimal model of self-propelled and self-aligning particles with nearest-neighbor elastic spring-like interactions on a fixed cell-cell connectivity graph~\cite{laang2024topology}. While there are no cell rearrangements or proliferation/extrusion events included in this active solid model, the interplay between elastic forces and self-aligning torques is sufficient to generate large density fluctuations acting as seeds for defect nucleation. 

Bacterial biofilms serve as a versatile experimental model for studying dense biological systems, where the active particles are embedded within an elastic network. A recent study~\cite{xu2023autonomous} shows that bacterial biofilms confined to a disk behave as an excitable medium, exhibiting self-sustained waves and emergent global
migration modes, such as i) oscillatory translation - synchronized back-and-forth motion, and ii) oscillatory rotation - global rotational motion with periodic chirality switches. To explain these collective behaviors, a minimal active elastic model was proposed and shown to reproduce main experimental observations. Their findings highlight that elastic restoring forces, self-propulsion and self-aligning torques are important minimal ingredients driving collective dynamics in dense active matter. 

An alternative formulation of self-aligning particles was proposed in Ref.~\cite{nissen2018theoretical} whereby the restoring torques act on intrinsic cell polarities, such as the PCP polarity and the apical-base polarity (AB), while these cell polarities mediate the pairwise forces between nearest-neighboring cells on a lattice. In a generic frame, the overdamped dynamics for the particle positions $\mathbf r_i$ and the corresponding unit vectors $\hat{\mathbf p}_i$ for PCP polarity and  $\hat{\mathbf n}_i$ for AB polarity read as  
\begin{eqnarray}
    \dot{\mathbf r}_i &=&\frac{1}{\gamma} \mathbf F_i+\sqrt{2D_t} \xi_i\\
    \dot{\hat{\mathbf p}}_i &=&(\mathbf T^{(p)}_i \times\hat{\mathbf p}_i) +\sqrt{2D_r} \eta_i \hat{\mathbf p}^\perp_i \\
    \dot{\hat{\mathbf n}}_i &=& (\mathbf T^{(n)}_i \times\hat{\mathbf n}_i) +\sqrt{2D_r} \chi_i \hat{\mathbf n}_i^\perp,
\end{eqnarray}
where the translational force $\mathbf F_i = -\sum_{j\ne i}\nabla_{\mathbf r_i} V_{ij}$, and aligning torques $\mathbf T^{(p)}_i = \sum_{j\ne i} \hat{\mathbf p}_i\times\nabla_{\hat{\mathbf p}_i} V_{ij}$ and $\mathbf T^{(n)}_i = \sum_{j\ne i} \hat{\mathbf n}_i\times\nabla_{\hat{\mathbf n}_i} V_{ij}$, are determined by a pairwise interaction potential $V_{ij}$, which couples the pair distance vector $\mathbf r_{ij}$ with local cell polarities. Based on general symmetry arguments of global invariance with respect to translations and rotations, the pairwise potential $V_{ij}$ for nearest neighbor interactions can be expressed in terms of bi-quadratic couplings as
\begin{equation}
V_{ij} = \lambda_1(\hat{\mathbf n}_i \times \hat{\mathbf r}_{ij}) \cdot (\hat{\mathbf n}_j \times \hat{\mathbf r}_{ij})+\lambda_2(\hat{\mathbf n}_i \times \hat{\mathbf p}_{j}) \cdot (\hat{\mathbf n}_j \times \hat{\mathbf p}_{i})+\lambda_3(\hat{\mathbf p}_i \times \hat{\mathbf r}_{ij}) \cdot (\hat{\mathbf p}_j \times \hat{\mathbf r}_{ij}).
\end{equation}
In Ref.~\cite{ho2024role}, it was shown that for a flat tissue in a disk confinement, the PCP polarities self-organize into spatial patterns determined by $+1$ defects. Topological ordering of cell polarities induces guided collective migration, i.e. inward aster motion for PCP vortex, outward spiral motion for PCP spirals and outward aster motion for PCP asters. The PCP vortex acts as source of inward migrations, which leads to out-of-plane migration and the formation of tubular structures~\cite{ho2024role}, also discussed in Sec.~\ref{Sec:biology}. There is an interesting interplay between topological defects within collective migration and the spontaneous cell neighbor exchanges leading to localized tissue fluidization, which needs to be further studied and understood.  

%---- dynamics of interacting topological defects -------
\section{Active turbulence in biological systems}
Active turbulence has been studied in bacterial motility, where dense suspensions of self-propelled bacteria - such as \textit{Bacillus subtilis} and \textit{Escherichia coli} - generate chaotic vortical flows. These flows arise due to  topological defects in the nematic alignment of bacteria suspension~\cite{dombrowski2004self, wensink2012meso, dunkel2013fluid, gachelin2014collective, peng2021imaging}. Unlike classical turbulence, which arises from inertial forces and the separation of energy injection and dissipation length scales, active turbulence is driven by the internal energy supplied at the single-cell level by swimming bacteria. This results in the formation of vortices and spectral energy transport at low Reynolds numbers~\cite{dombrowski2004self}. 

Beyond bacterial suspensions, active turbulence has been observed across a wide range of biological and synthetic systems, including reconstituted microtubule networks (where microtubule filaments interact with kinesin motor proteins)~\cite{opathalage2019self, giomi2015geometry, sanchez2012spontaneous, martinez2019selection}, sperm cell collectives~\cite{creppy2015turbulence}, and fluid-like epithelial layers undergoing jamming transitions ~\cite{blanch2018turbulent, lin2021energetics}. 

While active polar particles are intrinsically self-propelled, they can also align their principal elongation axis to form structural configurations with nematic (orientational) order. Often in this active nematics, there is no polar flow (no flocking migration), but instead there are dipolar flows generated by $\pm 1/2$ defects in the nematic alignment of elongated entities.  
Within these nematic phases, motile $+1/2$ defects generate dipolar flows with two bound and counter-rotating vortices, whereas $-1/2$ defects are associated with six bound flow vortices with alternating circulation \cite{yashunsky2024topological, yashunsky2022chiral, blanch2018turbulent, giomi2015geometry}. In confined geometries, nematic defects can drive emergent collective behaviors. For example, in human fibrosarcoma cell monolayers confined within microchannels, defect-laden turbulence develops persistent edge currents~\cite{yashunsky2022chiral}.

The concept of hyperuniformity, as a robust statistical measure of systems that approach uniformity at large scales despite appearing disordered at smaller scales, has been recently explored in the context of active turbulence. In bacterial turbulence, the density of $\pm 1/2$ defects exhibits hyperuniformity, characterized by suppressed long-range density fluctuations \cite{yashunsky2024topological}. It turns out that hyperuniform states also form in the distribution of flow micro-vortices and may correspond to optimal 
strategies of evasion and foraging motilities \cite{backofen2024nonequilibrium}. 

%--------------------------------

% Summary Points
\begin{summary}[SUMMARY POINTS]
\begin{enumerate}
\item Polar defects can be important for self-organization by mediating transitions between collective states such as flocking, milling, swirling, and turbulence in both biological and synthetic systems.

\item Polar defects act as organizing centers in various biological systems, influencing tissue morphogenesis, and large-scale cellular flows, with key examples found in the cornea epithelium, crypt-villi organization in the small intestine, and rotating \textit{Drosophila} egg chambers.

\item We can use experiments and theoretical models to explore defect-driven phase ordering and emergent structures, including stable asters, vortices, and dynamic polar textures in epithelial sheets, cytoskeletal filaments, and active colloidal systems.

\item Environmental factors such as confinement, curvature, and disorder strongly influence defect behavior, leading to distinct phenomena such as vortex-glass states, defect-driven flocking, and curvature-induced cell migration.
\end{enumerate}
\end{summary}

% Future Issue
\begin{issues}[FUTURE CHALLENGES]
\begin{enumerate}
\item Quantitative understanding of the interplay between topological defects and curvature of biological surfaces — such as in corneal, intestinal, and tubular epithelium — remains an open challenge. This requires studies on how defects shape cellular (polar or apolar) flows in three-dimensional environments.

\item The connection between topological defects and epithelial unjamming needs further studies, as defects may act as nucleation sites for collective migration, tissue remodeling, and symmetry breaking.

\item Bridging the gap between synthetic and biological active matter requires a deeper understanding of how mechanical forces, biochemical signaling, gene regulatory networks, and defect dynamics interact to regulate self-organization across different systems.

\item Developing a more unified theoretical framework for topological defects in active matter is essential, especially when extending to three dimensions, where defects and collective dynamics are intricately coupled with curved geometry.

\item Elucidate how defect ordering kinetics, coupled with cell density fluctuations, guide epithelial directional flow  in tissue repair.

\end{enumerate}
\end{issues}

%Disclosure
\section*{DISCLOSURE STATEMENT}
The authors are not aware of any affiliations, memberships, funding, or financial holdings that
might be perceived as affecting the objectivity of this review.  

%If the authors have nothing to disclose, the following statement will be used: The authors are not aware of any affiliations, memberships, funding, or financial holdings that
%might be perceived as affecting the objectivity of this review. 

% Acknowledgements
\section*{ACKNOWLEDGMENTS}
LA acknowledges support in part by grant NSF PHY-2309135 to the Kavli Institute for Theoretical Physics (KITP support from the KITP program Active Solids: From Metamaterials to Biological Tissue).

% References
%
% Margin notes within bibliography
%\section*{LITERATURE\ CITED}
\bibliographystyle{plainnat}  % or abbrvnat, unsrtnat
\bibliography{refs}

\end{document}